\documentclass[aps,prx,superscriptaddress,twocolumn]{revtex4}
\usepackage{amsfonts}
\usepackage{amsmath}
\usepackage{amsthm}
\usepackage{amscd}
\usepackage{amssymb}
\usepackage{subfigure}
\usepackage{amsxtra}
\usepackage{bm}           
\usepackage{bbm}
\usepackage{graphicx}
\usepackage{epstopdf}
\usepackage{color}
\usepackage{enumitem}
\usepackage{hyperref}

\def\bi#1\ei {\begin{itemize}#1\end{itemize}}
\def\bn#1\en {\begin{enumerate}#1\end{enumerate}}
\def\bea#1\eea {\begin{align}#1\end{align}}
\def\bean#1\eean {\begin{align*}#1\end{align*}}
\def\ben#1\een {\begin{equation*}#1\end{equation*}}
\def\be#1\ee {\begin{equation}#1\end{equation}}
\def\bes#1\ees {\begin{equation}\begin{split}#1\end{split}\end{equation}}
\def\bear#1\eear {\begin{eqnarray}#1\end{eqnarray}}
\def\bear#1\eear {\begin{eqnarray*}#1\end{eqnarray*}}

\newcommand{\beq}{\begin{equation}}
\newcommand{\eeq}{\end{equation}}

\newcommand{\ket}[1]{\ensuremath{\left|#1\right\rangle}}
\newcommand{\braket}[2]{\ensuremath{\langle #1|#2\rangle}}

\begin{document}
\title{Quantum superiority for verifying NP-complete problems with linear optics}
\author{Juan Miguel Arrazola}
\affiliation{Centre for Quantum Technologies, National University of Singapore, 3 Science Drive 2, Singapore 117543}
\author{Eleni Diamanti}
\affiliation{LIP6, CNRS, Universit\'{e} Pierre et Marie Curie, Sorbonne Universit\'{e}s, 75005 Paris, France}
\author{Iordanis Kerenidis}
\affiliation{IRIF, CNRS, Universit\'{e} Paris Diderot, Sorbonne Paris Cit\'{e}, 75013 Paris, France}
\affiliation{Centre for Quantum Technologies, National University of Singapore, 3 Science Drive 2, Singapore 117543}
\begin{abstract}
Demonstrating quantum superiority for some computational task will be a milestone for quantum technologies and would show that computational advantages are possible not only with a universal quantum computer but with simpler physical devices.
Linear optics is such a simpler but powerful platform where classically-hard information processing tasks, such as Boson Sampling, can be in principle implemented.
In this work, we study a fundamentally different type of computational task to achieve quantum superiority using linear optics, namely the task of verifying NP-complete problems. We focus on a protocol by Aaronson et al. (2008) that uses quantum proofs for verification. We show that the proof states can be implemented in terms of a single photon in an equal superposition over many optical modes. Similarly, the tests can be performed using linear-optical transformations consisting of a few operations: a global permutation of all modes, simple interferometers acting on at most four modes, and measurement using single-photon detectors. We also show that the protocol can tolerate experimental imperfections.
\end{abstract}
\date{\today}
\maketitle

\section{Introduction}

Quantum mechanics offers unprecedented possibilities to transmit and process information that have the potential to revolutionize information and communication technologies. While many such advantages are well understood theoretically, building a large-scale universal quantum computer or a fully-connected quantum internet remain formidable tasks for the not-so-near future. Towards these goals, it is important and worthwhile to identify examples where quantum superiority can be achieved using physical systems realizable with current or emerging technologies.

It has been fruitful to focus on specific physical systems and search for tasks that are well suited to be deployed in such platforms and where a quantum advantage can be demonstrated. A prime example of this is linear optics, namely the set of transformations on optical modes which preserve the total photon number. Linear optics can be used to perform universal quantum computing \cite{knill01a,browne04suba,kok2007linear,o2007optical}, increase the precision of estimation in metrology \cite{giovannetti2006quantum,giovannetti2011advances,toth2014quantum} and run efficient quantum protocols in communication complexity \cite{arrazolaqfp,arrazola2014QC,xu2015experimental,guan2016observation,kumar2017efficient}. On the road to achieving universal quantum computing, it has also become interesting to study specific tasks, notably Boson Sampling \cite{aaronson2011computational,broome2013photonic,tillmann2013experimental,crespi2013integrated,spring2012boson,spagnolo2014experimental}, where a computational advantage may be demonstrated by a linear optics scheme that is simpler to implement than a universal quantum computer.

Boson Sampling is a canonical example of a task suitable for quantum superiority. There, the task is to sample from the distribution that arises when a number of photons starting in some optical modes go through a circuit composed of beamsplitters and phase-shifters. While this task is in theory possible to perform by just running the corresponding linear optics circuit, it is related to some computationally hard problems in classical computation. There are other proposals for showing a quantum advantage in a computational context, including for example sparse commuting quantum circuits (IQP), where a randomly chosen IQP circuit is applied to a square lattice of $N$ qubits  \cite{boixo2016characterizing,farhi2016quantum,bremner2016achieving,bravyi2017quantum,gao2017quantum,bermejo2017architectures}. Note that both above mentioned examples perform circuits of depth at least $\sqrt{N}$.


The above proposals, however, present some drawbacks. First, the real difficulty in performing these tasks in a classical computer remains unclear, since it is based on unproven conjectures. In fact, recent results \cite{clifford2017classical,neville2017no}
provide much faster classical algorithms for Boson Sampling, implying that quantum superiority may need a system with a very large number of photons and optical modes. Second, while in theory we know what the linear optics system is supposed to do, one cannot verify whether the physical implementation actually works correctly or not. In other words, we have no means of testing if our linear optics system works as it should. Third, Boson Sampling or random IQP circuits do not correspond to problems required for real-world applications. Hence, finding an interesting computational task whose classical hardness is well established and which can be solved efficiently and in a verifiable way by a linear optics system remains a challenge.

Here, we deviate considerably from all previous examples and describe a fundamentally different type of computational task to achieve quantum superiority using linear optics. More precisely, we consider the task of verifying NP-complete problems, for example verifying whether a boolean formula is satisfiable or not.

In this setting, an all powerful but untrusted prover -- usually denoted as Merlin -- gives a witness of the solution, for example the truth assignment that satisfies the boolean formula, to an honest but computationally bounded verifier -- referred to as Arthur -- who checks the validity of the witness. By definition of the complexity class NP, if a witness exists, it is always possible for Merlin to provide a proof to Arthur who can verify it in polynomial time; for example Merlin can just provide the truth assignment that satisfies the boolean formula. If no witness exists, i.e. if the formula is unsatisfiable, then Arthur will always reject no matter what Merlin sends to him. But what happens if we restrict the amount of information that can be revealed to Arthur by these proofs? Can Arthur still verify if the formula is satisfied when he receives a proof that does not reveal much information about the satisfying truth assignment? In fact, if the revealed information is sufficiently small, it is in general no longer possible for Arthur to perform an efficient verification. Thus, in the case of verification where the revealed information is restricted, it might in principle be possible that quantum proofs can be verified more efficiently than classical ones, giving rise to a computational advantage. Indeed, it was shown in Ref. \cite{unentanglement} that for any NP-complete problem of size $N$, Merlin can send $O(\sqrt{N})$ quantum proofs each revealing $O(\log{N})$ bits of information, so that, under the promise that the proofs are not entangled with each other, they can be verified by Arthur in polynomial time on a quantum computer. On the other hand, in the classical case,  we will see that any verification algorithm acting on proofs that reveal at most $O(\sqrt{N}\log N)$ bits of information -- as in the quantum protocol -- must 
run in exponential time.

In this work, we show that the verification protocol of Ref.~\cite{unentanglement} for NP-complete problems can be implemented using simple linear-optical circuits and photonic sources. We assume, of course, that the prover has access to a classical witness when it exists. The proof states are implemented in terms of a single photon in an equal superposition over many optical modes, while the linear-optical transformations employed in the verification can be decomposed in terms of two main operations: a global permutation of all modes, and simple interferometers acting on at most four modes. As a consequence, the experimental requirements are significantly less stringent than those needed for linear-optics quantum computing or for performing arbitrary linear optics transformations. Our results illustrate another example of a computational quantum advantage in a linear optics setting. Moreover, we show that the protocol can tolerate experimental imperfections such as limited visibility and losses.

Let us make a few remarks about our result. First, the classical hardness of the problem is based on a well-established and widely believed conjecture, the Exponential Time Hypothesis \cite{impagliazzo99}, namely that the best classical algorithm for NP runs in time $2^{\delta N}$ for a constant $0 < \delta \leq 1$. In fact, the Strong Exponential Time Hypothesis \cite{dantsin10} claims that $\delta=1$. Second, the validity of the quantum circuit can be easily verified by running it on instances for which we already know the answer. Third, there is a vast number of NP-complete problems that arise naturally in all sciences and being able to verify them is an important task. Restricting the information leaked from the proofs is also a subject that has been extensively studied in the area of Zero Knowledge proofs \cite{goldwasser89} and it is relevant in cases where privacy is important. The fact that one can perform this verification with a simple linear optics system provides more evidence of the power and versatility of linear optics. Last, we note that our task is not solved by a typical quantum circuit, but involves an interaction between two parties; hence the quantum superiority is not for solving a computational task but for verifying efficiently the solution of a computationally hard problem.

The remainder of this paper is organized as follows. First, we review the verification protocol for the NP-complete problem 2-out-of-4 SAT of Ref.~\cite{unentanglement}, which consists of three tests that Arthur must be able to perform. By definition of NP, all other problems in the class can be reduced to 2-out-of-4 SAT with only a polynomial overhead and then verified. We describe how the proof states can be implemented and how each of these tests is carried out in a linear-optical setting. We conclude by analyzing the role of experimental imperfections in the protocol.

\section{Quantum Verification of 2-out-of-4 SAT}

In the 2-out-of-4 satisfiability problem (2-out-of-4 SAT), we are given a formula over $N$ binary variables consisting of a conjunction of clauses, each of which contains exactly four variables. The clauses are satisfied if and only if exactly two variables are equal to 1, i.e. if $x_i+x_j+x_k+x_l=2$ for a clause relating the variables $x_i,x_j,x_k$ and $x_l$. The problem is to decide whether there exists an assignment $x=x_1x_2\cdots x_N$ such that the formula is satisfied. We focus on the case in which the 2-out-of-4 SAT instance meets two conditions. First, the instance must be balanced, meaning that every variable occurs in at most a constant number of clauses, and furthermore, the instance must be a PCP, i.e. either it is satisfiable or else a fraction of at most $1-\epsilon$ of the clauses should be satisfiable, for some $\epsilon>0$. Note that these conditions can always be guaranteed when reducing the 3-Satisfiability (3SAT) problem to 2-out-of-4 SAT \cite{dinur, unentanglement}, and therefore any NP-complete problem can be reduced to an instance of 2-out-of-4 SAT satisfying these restrictions by first reducing it to an instance of 3SAT.

In a valid verification protocol, if there exists a satisfying assignment for the instance, then a correct proof is accepted by Arthur with high probability -- typically larger than $2/3$. This property is called completeness. Similarly, if there is no satisfying assignment for the instance, then any proof is rejected by Arthur with high probability -- again, typically larger than $2/3$. This property is known as soundness. In Ref. \cite{unentanglement}, it was shown that there exists a quantum verification protocol for 2-out-of-4 SAT that is both sound and complete. In the following, we describe how this protocol can be carried out in a linear-optical setting.

\subsection{State preparation}

Since we are verifying NP-complete problems, we have to assume that the prover has access to the classical witness, otherwise there would be an efficient algorithm for NP, which is highly unlikely. Then, the first ingredient in the verification protocol is the construction of the quantum proofs. Merlin sends Arthur $K$ proofs $\ket{\psi_x}_1\otimes\ket{\psi_x}_2\otimes\ldots\otimes\ket{\psi_x}_K$, with $K=O(\sqrt{N})$. Each of these proofs is an $N$-dimensional state of the form
\beq
\ket{\psi_x}=\frac{1}{\sqrt{N}}\sum_{i=1}^N(-1)^{x_i}\ket{i},
\eeq
where $x$ is the string satisfying the instance of the 2-out-of-4 SAT problem. We henceforth refer to any state of this form as a proper state.

Note that this state is mathematically equivalent to a state of $\lceil \log_2 N \rceil$ qubits and therefore it can only reveal at most $\lceil \log_2 N \rceil$ bits of information about $x$.  In a linear-optical setting, this state can be implemented in terms of a single photon in a superposition over $N$ different modes as
\beq\label{EQ:properstate}
\ket{\psi_x}=\frac{1}{\sqrt{N}}\sum_{i=1}^N(-1)^{x_i}a^{\dagger}_i\ket{0},
\eeq
where $a^{\dagger}_i$ is the creation operator for the $i$-th mode.

The soundness proofs of Ref.~\cite{unentanglement} assume that Merlin can only send states in a Hilbert space of dimension $N$, which in this case corresponds to the single-photon subspace of the $N$ modes. To ensure soundness of the verification, Arthur simply rejects the proof if he observes more than one photon in the states he measures. Indeed, in this case, for any strategy in which Merlin sends states containing $n$ photons with probability $P(n)$, the acceptance probability by Arthur obeys
\begin{align}
P(\textrm{accept})=&\sum_{n=0}^{\infty}P(\textrm{accept}|n)P(n)\nonumber\\
=&P(\textrm{accept}|1)P(1)\label{P(1)}
\end{align}
since $P(\textrm{accept}|n)=0$ for any $n\neq 1$. This probability is maximized for $P(1)=1$ and therefore we ensure that Merlin's optimal strategy employs single photon states, in which case the soundness proof of Ref. \cite{unentanglement} holds.

It is also required that the states are not entangled with each other. There is no known general method of detecting this entanglement \cite{unentanglement} and therefore Arthur cannot enforce this condition through a test in his verification. We thus view this unentanglement condition as a promise on the form of the proofs produced by Merlin. Note that this requirement can be enforced if we assume that Arthur interacts with $\sqrt{N}$ non-communicating provers that do not share any entanglement \cite{unentanglement}. In fact, the unentanglement condition can be enforced even if there are just two non-communicating provers that do not share any entanglement \cite{2provers}.


One way for Merlin to create the state $\ket{\psi_x}$ is to start with an initial state of the form
\beq\label{EQ:EqualSupStates}
\ket{\psi}=\frac{1}{\sqrt{N}}\sum_{i=1}^N a^{\dagger}_i\ket{0},
\eeq
and then have $N$ phase-shifters acting on each of the $N$ modes which apply a phase-shift of $-1$ only when the corresponding bit of the classical witness $x$ is 1. This allows Merlin to perform the transformation $\ket{\psi}\rightarrow \ket{\psi_x}$ as desired.

\begin{figure}[t!]
\begin{center}
\includegraphics[width=0.8\columnwidth]{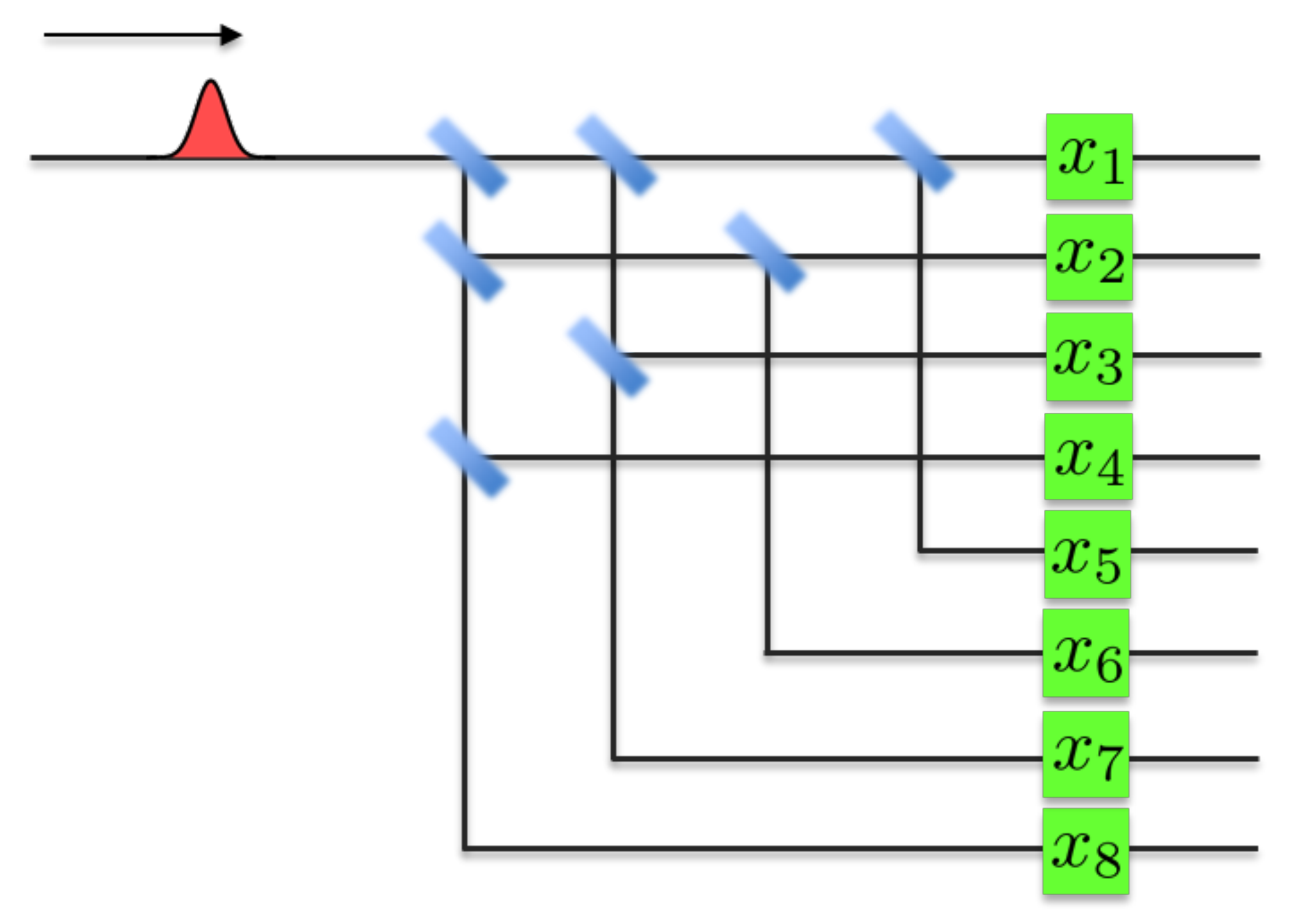}
\caption{(Color online) Circuit for creating the proof states used in the verification protocol, illustrated for $N=8$. A single photon passes through a cascade of beamsplitters to create an equal superposition state over all the output modes, which are then subject to a phase-shift which depends on the string $x$. }\label{Fig:StatePrep}
\end{center}
\end{figure}

Equal superposition states of the form of Eq. \eqref{EQ:EqualSupStates} can be created by sending a single photon through a cascade of beamsplitters (see Fig.~\ref{Fig:StatePrep}). An equal superposition state over $N$ modes can be implemented in this way using $O(\log N)$ beamsplitters resulting in a linear optics circuit of depth $O(\log N)$. Such circuits have been implemented for small $N$ \cite{heilmann2015novel}. The output modes are then sent through phase-shifters to create the proof states.

Once the proof states have been prepared, Arthur performs his verification which employs three tests: the satisfiability, uniformity, and symmetry tests. Arthur selects one of the three tests uniformly at random and decides whether to accept or reject the proof depending on the specific criterion of each test. Below we describe how these tests can be performed in a linear optics setting.

\subsection{Satisfiability test} 
\begin{figure}[h!]
\begin{center}
\includegraphics[width=0.75\columnwidth]{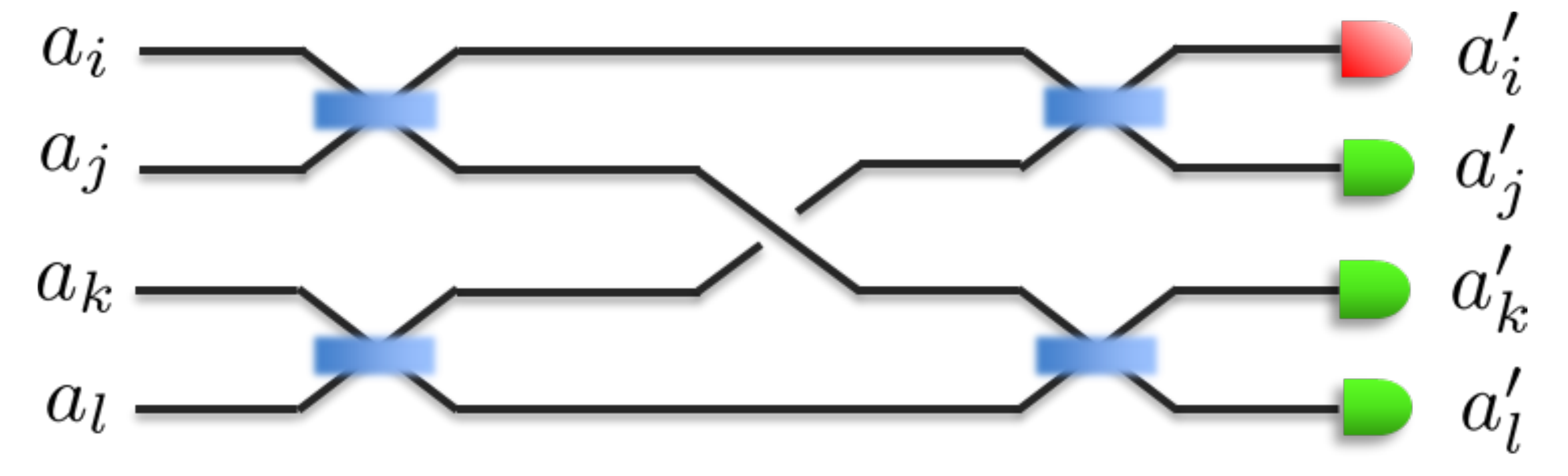}
\caption{(Color online) Interferometer for testing satisfiability of 2-out-of-4 SAT. In the ideal case, the detector for mode $a$ will never detect a photon if the clause is satisfied and there will be one photon detected in one of the other three modes with certainty. If the clause is not satisfied, there is at least a probability
$\Omega(1/N)$ of observing a photon in this mode.
}\label{Fig:circuit}
\end{center}
\end{figure}

In the satisfiability test, Arthur checks that the assignment $x$ which is encoded in the quantum proofs satisfies the 2-out-of-4 SAT instance. To do so, Arthur divides all the clauses into a constant number of blocks $B_1,B_2,\ldots,B_s$ in such a way that each block contains at least $\Omega(N)$ clauses and in each block, no variable appears more than once. This partition into different blocks is guaranteed to exist because the instance is balanced. Arthur selects a block {$B_r$} uniformly at random from this set. He then picks a state $\ket{\psi_x}$ at random from the $K$ copies and performs a permutation $\Pi_{Sat}$ of the modes that groups them into the clauses corresponding to the selected block {$B_r$}. After the permutation, for every clause of the form $x_i+x_j+x_k+x_l$, the corresponding modes $a_i,a_j,a_k,a_l$ have been placed in sequence beside each other. For each such set of four modes, Arthur interferes them in the circuit shown in Fig. \ref{Fig:circuit}.

The effect of this interferometer is to perform a mode transformation between the input modes $a_i,a_j,a_k,a_l$ and the output modes $a'_i,a'_j,a'_k,a'_l$ given by
\begin{align}
a_i&\longrightarrow\frac{1}{2}(a'_i+a'_j+a'_k+a'_l)\nonumber\\
a_j&\longrightarrow\frac{1}{2}(a'_i+a'_j-a'_k-a'_l)\nonumber\\
a_k&\longrightarrow\frac{1}{2}(a'_i-a'_j+a'_k-a'_l)\nonumber\\
a_l&\longrightarrow\frac{1}{2}(a'_i-a'_j-a'_k+a'_l)\nonumber.
\end{align}
For a single-photon proper state, as in Eq. \eqref{EQ:properstate}, this relation implies that the probability of observing a photon in each of the output modes is given by
\begin{align}
P_{a'_i}&=\frac{1}{4N}[(-1)^{x_i}+(-1)^{x_j}+(-1)^{x_k}+(-1)^{x_l}]^2\nonumber\\
P_{a'_j}&=\frac{1}{4N}[(-1)^{x_i}+(-1)^{x_j}-(-1)^{x_k}-(-1)^{x_l}]^2\nonumber\\
P_{a'_k}&=\frac{1}{4N}[(-1)^{x_i}-(-1)^{x_j}+(-1)^{x_k}-(-1)^{x_l}]^2\nonumber\\
P_{a'_l}&=\frac{1}{4N}[(-1)^{x_i}-(-1)^{x_j}-(-1)^{x_k}+(-1)^{x_l}]^2\nonumber.
\end{align}
Whenever the clause is satisfied, i.e. when $x_i+x_j+x_k+x_l=2$, a photon will never be detected in mode $a'_i$. We refer to this mode as the satisfiability mode. If the clause is not satisfied, the probability of observing a photon in the satisfiability mode is twice the one in the other three modes.

\begin{figure}[t!]
\begin{center}
\includegraphics[width=\columnwidth]{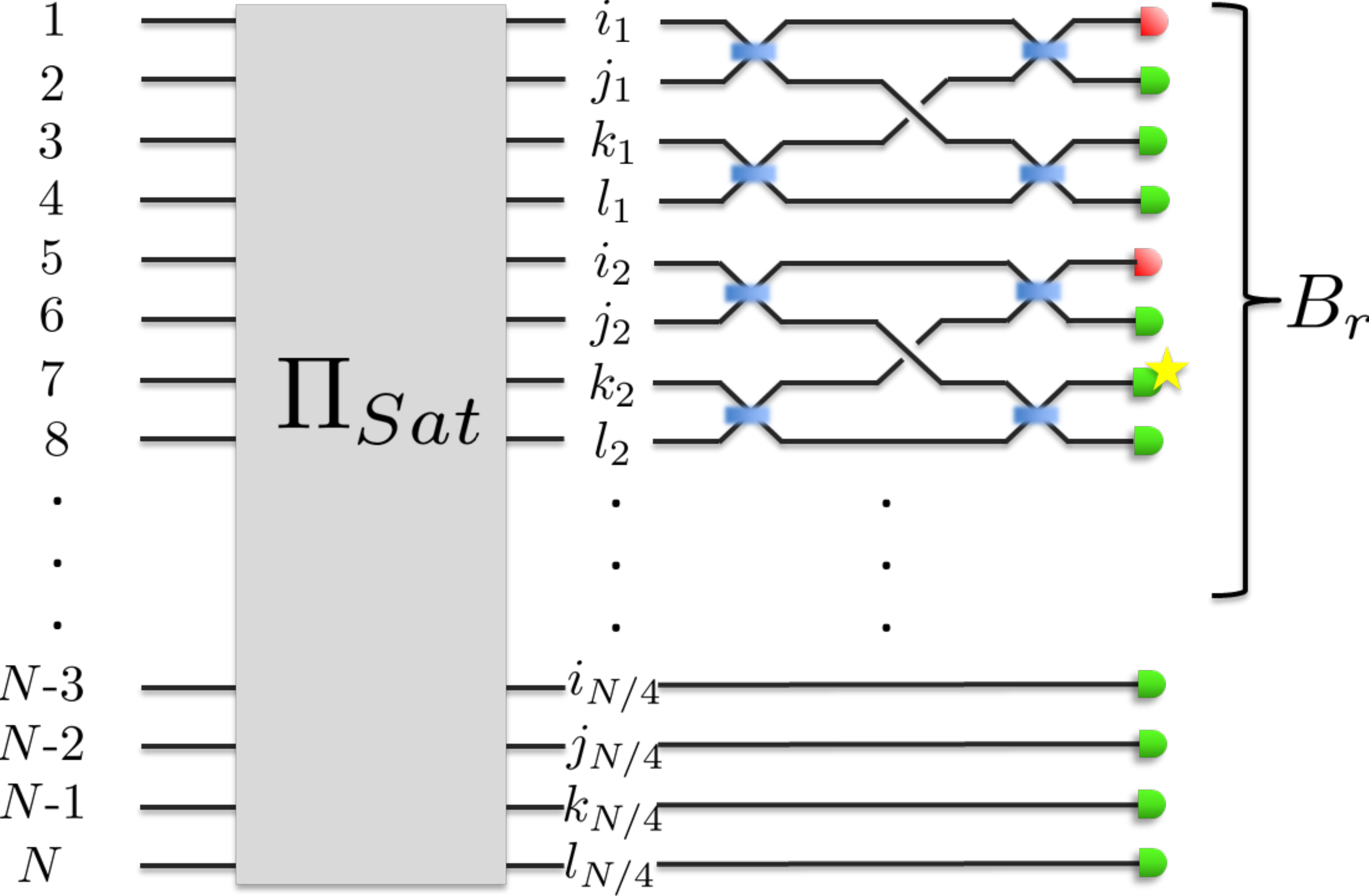}
\caption{(Color online) Linear-optical circuit for the satisfiability test. Arthur selects a block $B_r$ at random and performs a permutation $\Pi_{Sat}$ of the incoming modes that groups them according to the clauses in the block. The modes {of each clause} are then sent through a four-mode interferometer and Arthur checks for photons in the outputs. Modes not in the block are not interfered, but detection still takes place. He accepts the proof if at most one photon is detected and it does not happen in any of the satisfiability modes, which are depicted in red. }\label{Fig:SATInterferometer}
\end{center}
\end{figure}

Arthur's criterion for acceptance is the following: he accepts the proof if and only if exactly one photon is detected and it is not detected in a satisfiability mode. This is illustrated in Fig. \ref{Fig:SATInterferometer}. In the honest Merlin case, the test will pass with certainty while, as shown in Ref. \cite{unentanglement}, if $x$ is not a satisfying assignment of the problem, a constant fraction of the clauses will be unsatisfied, leading to an overall constant probability of rejecting the proof, since there are $\Omega(N)$ clauses in each block. Thus, the test has perfect completeness and constant soundness.

\subsection{Uniformity test} Arthur's satisfiability test functions correctly whenever the states sent by Merlin are proper states, i.e. of the form of Eq.~\eqref{EQ:properstate}. Arthur requires an additional test to certify that the states he receives are proper states. To perform this uniformity test, Arthur first selects a random perfect matching on the set $\{1,2,\ldots,N\}$. A perfect matching is a partitioning of the set into $N/2$ disjoint edges $\{(i_1,j_1),(i_2,j_2),\ldots,(i_{N/2},j_{N/2})\}$. For instance, $\{(1,3),(2,5),(4,6)\}$ is a possible matching on the set $\{1,2,3,4,5,6\}$. For each of the $K$ states he receives, Arthur performs a permutation of the modes such that all modes are paired according to the edges $(i,j)$ in the matching. After the permutation, for every edge he interferes the corresponding pair of modes in a 50:50 beam-splitter and checks for photons in the outputs. The beam-splitter performs the transformation
\begin{align}
a_i&\longrightarrow\frac{1}{\sqrt{2}}(a'_i+a'_j),\hspace{0.5cm}a_j\longrightarrow\frac{1}{\sqrt{2}}(a'_i-a'_j)\nonumber.
\end{align}
which means that, for a proper state, the probabilities of observing a photon in each output are
\begin{align}
P_{a'_i}&=\frac{1}{2N}[(-1)^{x_i}+(-1)^{x_j}]^2\nonumber\\
P_{a'_j}&=\frac{1}{2N}[(-1)^{x_i}-(-1)^{x_j}]^2\nonumber.
\end{align}
Thus, whenever a photon is detected in a pair of modes $(a'_i,a'_j)$, Arthur learns the value of  $x_i\oplus x_j$. This allows a labelling of all possible outcomes of this measurement as $(i,j,b)$, with $b=x_i\oplus x_j$.

\begin{figure}[t!]
\begin{center}
\includegraphics[width=0.95\columnwidth]{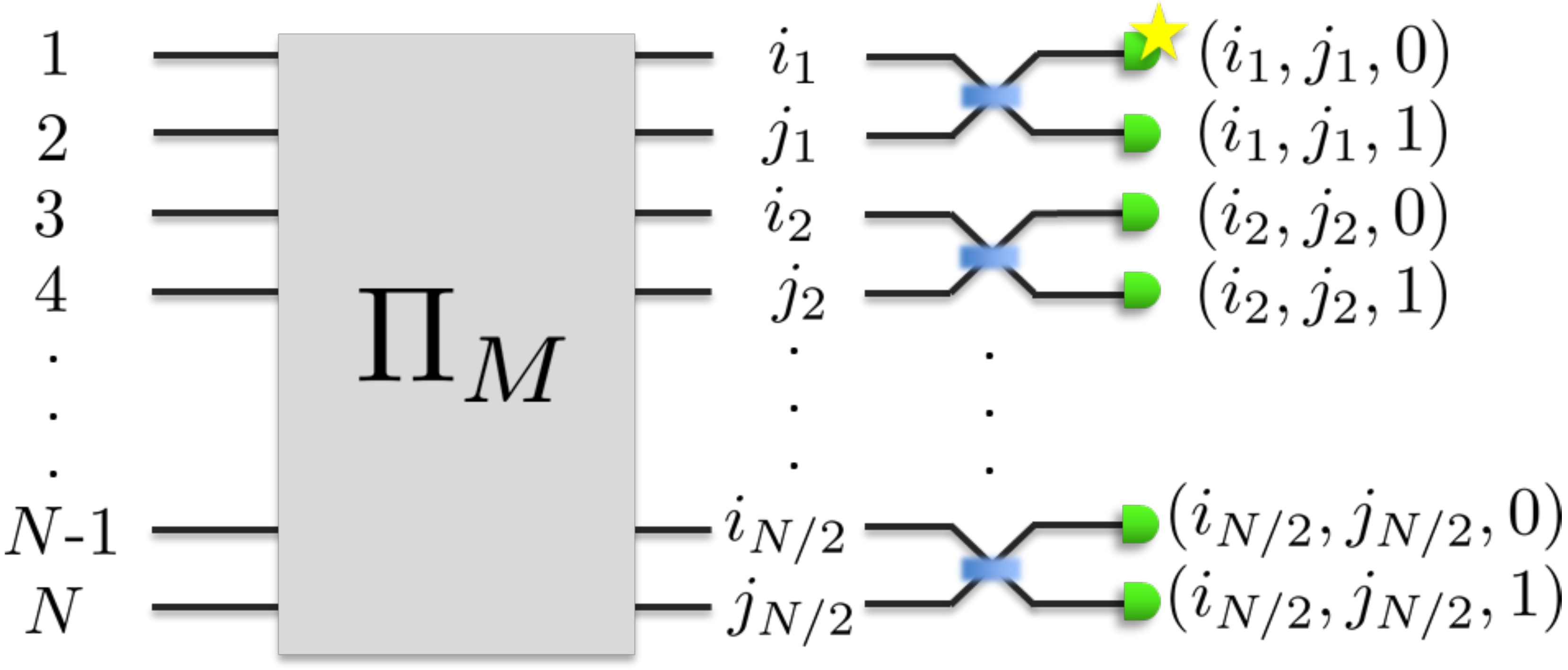}
\caption{(Color online) {Linear-optical circuit for the uniformity test for one state}. Arthur selects a matching at random and performs a permutation $\Pi_M$ of the $N$ modes that pairs them according to the edges of the matching. The pairs of modes interfere in 50:50 beamsplitters and Arthur checks for photons in the outputs. This circuit is used for each of the $K$ states and Arthur rejects the proof if he observes incompatible outcomes of the form $(i,j,0)$ and $(i,j,1)$ across different states. }\label{Fig:Uniformity}
\end{center}
\end{figure}

Arthur's uniformity test is the following: he performs the measurement described above on all $K$ copies of the state and he accepts the proof only if there are no incompatible outcomes of the form $(i,j,0)$ and $(i,j,1)$. As before, he also requires that there is exactly one photon detected in each state and he rejects the proof if there are no collisions, i.e. photons detected for the same edge $(i,j)$ in different copies. The test is illustrated in Fig. \ref{Fig:Uniformity}.

By choosing $K=O(\sqrt{N})$, it follows from the generalized birthday paradox \cite{knight1973birthday} that collisions will occur with high probability. In the honest case, incompatible outcomes never occur and therefore the test has constant completeness where rejection of a correct proof only occurs if there are no collisions. On the other hand, if the states are far from a proper state, it was shown in Ref. \cite{unentanglement} that the test has a constant probability of rejecting the proof for any other input state and thus has constant soundness.

\subsection{Symmetry test} If Arthur performs a satisfiability and a uniformity test, the only room left for Merlin to deviate from honest behaviour is to send different proper states in each of the $K$ systems. To protect against this, Arthur needs to check that all states are equal. He can achieve this by using a SWAP test: a two-outcome measurement on a pair of states with the property that the probability of obtaining each outcome depends on the inner product of the states.

\begin{figure}[t!]
\begin{center}
\includegraphics[width=0.75\columnwidth]{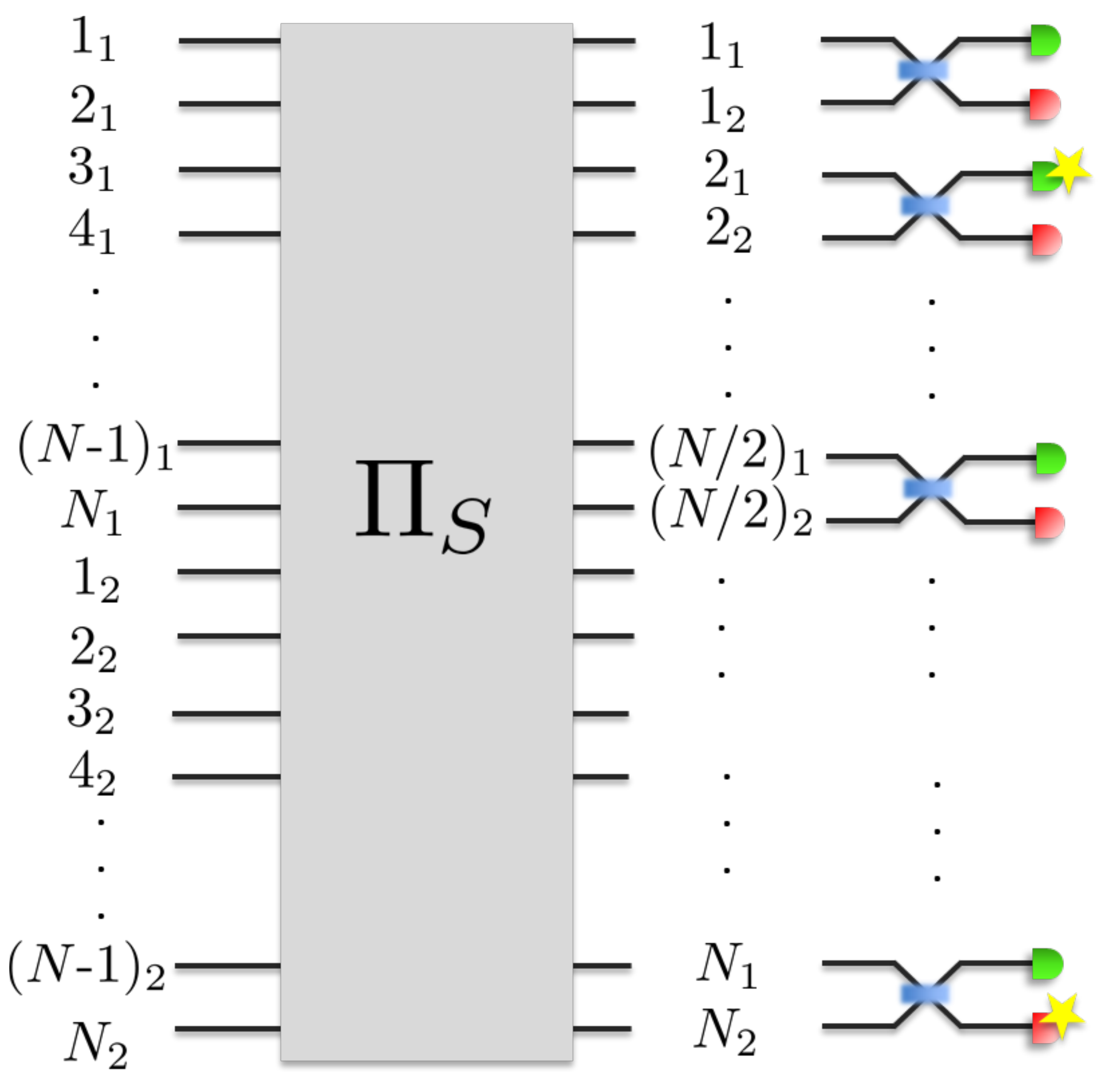}
\caption{(Color online) Arthur selects two out of the $K$ states uniformly at random and performs a permutation $\Pi_S$ that pairs the modes from each state, which are subsequently interfered in 50:50 beamsplitters. The outputs can be divided into a set of ``up" modes (depicted by red detectors) and ``down" modes (green detectors). The probability of observing a coincidence, i.e. a photon in an ``up" mode and the other photon in a ``down" mode, depends on the inner product of the states and never occurs if the states are equal. Arthur can use this property to detect if the states are different. This is shown in the figure by clicks occurring in an ``up" and a ``down" mode, which cause Arthur to reject.}\label{Fig:Symmetry}
\end{center}
\end{figure}

To perform the test in a linear optics setting, Arthur randomly selects two out of the $K$ states and performs a permutation that pairs the $i$-th mode of the first state with the $i$-th mode of the second  state for all $i=1,2,\ldots,N$. Afterwards, each pair of modes is sent through a 50:50 beamsplitter. The output modes of each beam-splitter can be labelled as the ``up" mode and the ``down" mode. The interference of the two photons corresponds to a generalized version of the Hong-Ou-Mandel effect \cite{hong1987measurement} and indeed it was shown in Ref.~\cite{garcia2013swap} that for any two proper input states $\ket{\psi}$ and $\ket{\phi}$ as in Eq. \eqref{EQ:properstate}, the probability of observing a coincidence, i.e. a photon in an ``up" mode and the other photon in a ``down" mode, is equal to $(1-|\braket{\psi}{\phi}|^2)/2$, the same probability of {a SWAP test resulting in a `different state' outcome.} 
In particular, this implies that coincidences never occur if the states are equal.

This property allows Arthur to perform the following test for symmetry: he accepts the proof if and only if exactly two photons are detected and there are no coincidences. Note that the SWAP test is a crucial component in quantum fingerprinting \cite{QuantumFingerprinting,arrazolaqfp} and its implementation in a linear-optical setting has already been demonstrated in recent quantum fingerprinting experiments \cite{xu2015experimental,guan2016observation}. The symmetry test is illustrated in Fig. \ref{Fig:Symmetry}.

In the honest case, the test passes with certainty and therefore it has perfect completeness, while it was shown in Ref.~\cite{unentanglement} that the test also has constant soundness. Overall, by selecting randomly between these three tests, Arthur can verify Merlin's proof with perfect completeness and constant soundness, as required by a verification protocol.\\

For the {satisfiability and symmetry} tests described above, the verifier needs to randomly pick one or a pair of the $K$ proofs on which to apply the test. Note that the verifier can pick this before the proofs arrive. One way of doing this is a $K \times K$ block switch that takes as input the $K$ proofs, puts two random ones as the first two and leaves the remaining unchanged.

\begin{figure}[t!]
\begin{center}
\includegraphics[width=0.95\columnwidth]{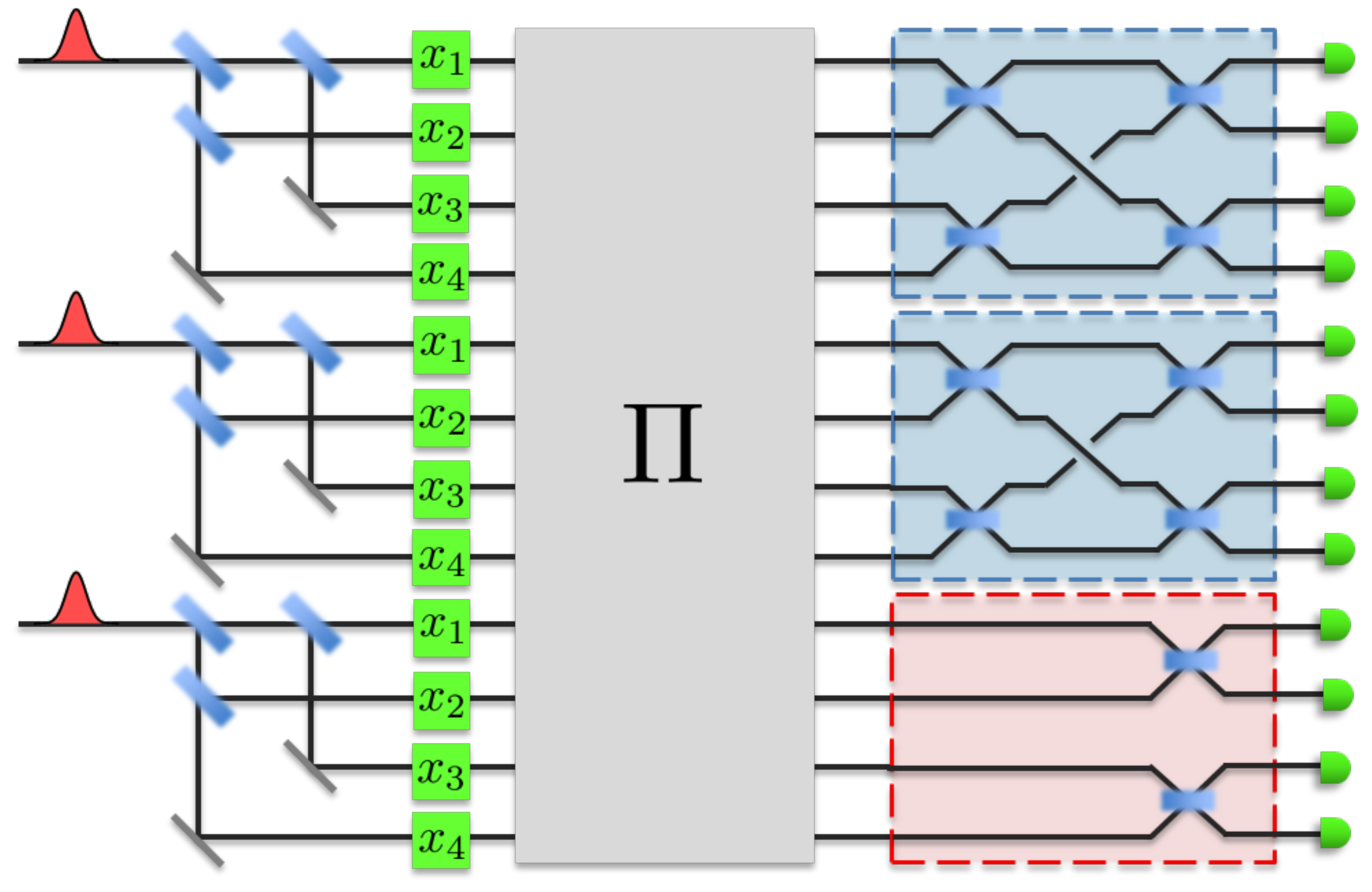}
\caption{(Color online). Complete setup for the linear-optical verification of 2-out-of-4 SAT. For illustration, we consider the case of $N=4$ and $K=3$. Merlin prepares three equal-superposition states of a single photon. Each mode passes through phase shifters (green) to encode the satisfying assignment $x$ as in Eq. \eqref{EQ:properstate}. Arthur then applies a permutation $\Pi$ on the modes depending on which of the three tests he is going to perform and which proofs he randomly picked. At the output of the permutation, he attaches either four-mode interferometers for the satisfiability test (blue), or 50:50 beamsplitters to each pair of modes (red) for the other two tests. He checks for photons in the outputs and decides whether to accept or reject depending on the pattern of clicks observed.}\label{Fig:Full}
\end{center}
\end{figure}

In summary, to perform the verification of 2-out-of-4 SAT for an instance of size $N$, Arthur needs the following components when using spatial modes:
\begin{enumerate}[noitemsep]
\item $K=O(\sqrt{N})$ single photon sources.
\item $K$ fixed cascades of beamsplitters of depth $O(\log N)$, each {preparing} 
a single photon in an equal superposition over $N$ modes.
\item $KN$ phase-shifters, one for each mode.
\item  One $K \times K$ block switch that permutes groups of $N$ modes.
\item $K$ $N \times N$ switches that perform arbitrary permutations of $N$ modes.
\item One $2N \times 2N$ switch that performs an arbitrary permutation of $2N$ modes.
\item $O(N)$ four-mode interferometers for the satisfiability test.
\item $O(KN)$ two-mode interferometers for the uniformity and symmetry tests. {This includes the photon number resolving detectors.}
\end{enumerate}
A complete setup for the verification protocol is illustrated in Fig. \ref{Fig:Full}. Note that by separating the modes in time it would be possible to use a constant number of interferometers and detectors, greatly reducing the number of required components. {It is also possible to optimize the required resources by using delay circuits to suitably direct the selected proofs depending on the chosen test.}

In the next section, we calculate the running time of this quantum verification protocol showing that it runs in polynomial time. Note also that only $O(\sqrt{N} \log N)$ bits of information are revealed to the verifier about the witness $x$. Then, we show that any classical algorithm using proofs that reveal only $O(\sqrt{N} \log N)$ bits of information requires exponential time under the only assumption that there are no classical algorithms for NP-complete problems running in time less than $2^{O(N)}$.

\subsection{Running time of the verification algorithm}

The quantum verification procedure can be decomposed into three main steps: the preparation by the prover of the quantum states that correspond to the classical witness; a permutation circuit that the verifier uses to rearrange the optical modes according to his random choices; and interferometers running on at most four modes each. As discussed before, given knowledge of the classical witness $x$, each proof state can be prepared using a simple cascade of beamsplitters of size $O(N)$ and of depth $O(\log N)$ as well as phase-shifters, while the interference circuits for the tests have constant depth and $O(N)$ size. Let us look a bit more carefully at the permutation circuit which just spatially rearranges the modes. First, notice that all the random choices of the verifier can be made before he receives the proofs, so the entire permutation circuit can be prepared in advance. For the satisfiability and uniformity tests, we need a permutation module acting on $N$ modes, while in the symmetry test, the permutation acts on $2N$ modes. Such permutations can be performed using a universal circuit of size $O(N^2)$ \cite{reck94a,carolan2015universal}, but there also exist standard microelectromechanical system switches (MEMS) that perform such permutations using only $O(N)$ adjustable mirrors. MEMS switches have been demonstrated to work for up to 1100 modes \cite{kim20031100}. We also need a permutation to choose one or two of the $K$ proofs {for the satisfiability and symmetry tests}, which can also be performed by a switch of size  $O(K)=O(\sqrt{N})$. Hence, if the proofs arrive at the same time in different spatial modes, the size of the quantum circuit is $O(KN)$ (dominated by the state preparation).


For the running time of the quantum verification algorithm, let us consider first the case of all proofs arriving at the same time in different spatial modes.
Then, we divide the algorithm into two steps: the preprocessing of the permutation circuit, which takes time $O(N)$; and the quantum execution of the algorithm that takes time $O(\log N)$, which is the depth of the quantum circuit. Note that, using standard Chernhoff bound arguments, the verification error can be reduced to any small constant by simply repeating the protocol a constant number of times. In the case each proof comes sequentially, then the running time of the quantum execution is also $O(N)$ since we possibly need to wait until the last proof.

The states employed are each of dimension $N$, so the global state consisting of all copies has dimension $\log N^K= \gamma \sqrt{N}\log N$, for some small constant $\gamma$. The dimension of this state places an upper bound on the information that it contains about the classical witness $x$. In particular, for any proof state $\rho$, we have that the mutual information with the string $x$ satisfies
\beq\label{EQ:min-entropy}
I(X : \rho) = H(\rho) - H(\rho | X) \leq H(\rho)  \leq \gamma \sqrt{N}\log N.
\eeq
In summary, we have a quantum verification procedure where the verifier takes as input a proof revealing at most $ \gamma \sqrt{N}\log N$ bits of information about $x$ and can verify instances of 2-out-of-4-SAT in $O(N)$ time.

What can we say about the running time of a classical verification algorithm receiving a proof that reveals the same amount of information? Denote by $R(A)$ the running time of any verification algorithm acting on a classical proof $\phi$ with  $I (X : \phi) \leq H(\phi) \leq \gamma \sqrt{N}\log N$. We want to bound the asymptotic scaling of $R(A)$ by using the verification algorithm $A$ to produce an algorithm for 2-out-of-4 SAT by randomly searching over all possible proofs. The algorithm is the following:
\begin{enumerate}[noitemsep]
\item Generate a random proof. From the entropy bound, the probability of guessing a correct proof is $p_{guess}(\phi)  = 2^{-H_{min}(\phi)} \geq  2^{-H(\phi)}  \geq 2^{ -\gamma \sqrt{N}\log N}$.
\item Repeat the verification protocol on this proof $O(\sqrt{N}\log N)$ times and take the majority vote of the outcomes. This identifies whether or not the proof is correct except with probability $O(2^{-\gamma \sqrt{N}\log N})$.
\item Repeat the previous steps $O(2^{\gamma \sqrt{N}\log N})$ times to ensure that, if it exists, a correct proof will be found with high probability. Accept if a valid proof is found, reject otherwise.
\end{enumerate}

This is an algorithm for 2-out-of-4-SAT with running time $O(\sqrt{N}\log N 2^{ \gamma \sqrt{N}\log N})R(A)$. Under the Exponential Time Hypothesis that algorithms for NP-complete problems must have running time $2^{\delta N}$, we have that
\beq
O(\sqrt{N}\log N 2^{\gamma \sqrt{N}\log N})R(A)\geq 2^{\delta N}
\eeq
and therefore
\beq
R(A)\geq O\left(\frac{1}{\sqrt{N}\log N } 2^{\delta N - \gamma \sqrt{N}\log N}\right),
\eeq
which is exponential in the input size $N$ for large enough $N$. Since we showed that the quantum verification runs in polynomial time, we conclude that there is a quantum advantage in the linear-optical verification of an NP-complete problem with proofs revealing a restricted amount of information.

Let us now make a {quick} 
calculation to understand what order of $N$ we would possibly need in order to show quantum superiority. Note that the constant $\gamma$ comes from the uniformity test, where we need enough proofs to find a collision with high enough probability. Let us take this to be $\gamma = 2$. Let us also assume the Strong Exponential Time Hypothesis, namely $\delta=1$. In this case, we need to make sure that a classical algorithm which runs in time exponential in $N - 2\sqrt{N} \log N$ remains infeasible. By taking $N=512$, we have that the classical algorithm must run for time more than $2^{100}$. Note that $N$ is the number of optical modes for each of the $2\sqrt{N}$ proofs and that each proof contains one photon. In other words, our circuit has a total of 46 photons, each one in 512 optical modes. The number of photons is comparable to the other proposals, namely Boson Sampling and IQP circuits. While the number of optical modes in our scheme is significantly larger, the depth of the circuits is only logarithmic and not polynomial in $N$ as in the other proposals. This is of course a high level calculation and there is a lot of space for optimizing these parameters. On the other hand it is important to consider experimental imperfections as well.

In the following, we discuss the role of experimental imperfections in the quantum verification scheme, showing that they can be tolerated by 
increasing the number of copies by a constant factor.

\section{Experimental imperfections}
In linear optics, there are three main forms of experimental imperfections: detector dark counts, limited interferometric visibility, and losses. Let us start with dark counts. For a state of $N$ modes, where the dark count probability for each detector is $p_{dark}$, the probability of obtaining a {single click due to a dark count} is
\beq
{p_{click}= 1 - (1-p_{dark})^N\approx N p_{dark},}
\eeq
which is negligible as long as $N\ll 1/p_{dark}$. Typical values of the dark count probability are below $10^{-6}$ whereas, as discussed, a quantum advantage can be reached for values of $N$ many orders of magnitude smaller than $10^6$.

Limited interferometric visibility refers to all deviations from the ideal state preparation and transformations. This will not lead to a change in the expected number of clicks, but it can cause the wrong detectors to fire. The verification protocol can then tolerate limited visibility as long as there remains a constant gap between soundness and completeness of the test, {i.e., of the difference between the probability of accepting a correct proof and the one of accepting an incorrect one,} since this difference can be amplified by repeating the verification a constant number of times. Note that our protocol provides such a constant gap in the case of no imperfections.

Losses in the verification are problematic for the tests as we have previously defined them, since proofs are rejected if no photons are detected. To address this, we can modify the tests to correct for this effect. Let $\eta$ be the overall transmissivity of the protocol, meaning that a single photon is detected with probability $\eta$. We address the modifications to each test separately.

As stated previously, the satisfiability test acts on a single randomly chosen state and rejects if no photons are detected. In the presence of losses, this would cause the test to reject with probability $1-\eta$ even for a correct proof. Instead, we modify the test by instead randomly selecting $O(1/\eta)$ states to ensure a high probability of observing a photon in at least one state, and then performing the satisfiability test on each of them. The proof is then accepted if and only if no photons are detected in the satisfiability modes, not more than one photon is detected in each state, and there is at least one state for which a photon is detected.

Similarly, to test for symmetry, instead of randomly selecting a single pair of states, we must now randomly select $O(1/\eta^2)$ pairs of states to ensure a high probability of having two photons in at least one pair of states, and perform the symmetry test on each pair. In this case, we accept the proof if and only if no coincidences are observed for any state, no more than two photons are detected for any pair, and there is at least one pair for which two photons are detected.

Finally, in the uniformity test, a measurement is made on all $K=O(\sqrt{N})$ states, so in this case we can compensate for the presence of losses by increasing the number of copies to $K=O(\sqrt{N}/\eta)$ in order to ensure a high probability of obtaining a collision. Arthur accepts the proof if and only if no incompatible outcomes of the form $(i,j,0)$ and $(i,j,1)$ occur, at least one collision occurs, and not more than one photon is detected in any state.

Overall, these modifications to the tests lead to completeness of the verification in the presence of losses. With respect to the soundness, note that any statistics that Merlin can induce in Arthur's measurement in the presence of losses can also be obtained in the ideal lossless case, since Merlin can just introduce the losses himself -- so we recover the soundness of the lossless case.

Note also that what we want to demonstrate is a quantum circuit that verifies the NP-complete problem correctly and from which the verifier does not get more than  $O(\sqrt{N}\log N)$ bits of information about the classical witness. In practice one can ensure this by just making sure that the number of detector clicks are bounded, since this is the way the verifier obtains information. It is important to remember that we are not in a cryptographic setting where we have to worry about a verifier trying to cheat by changing the circuit in order to get more information from the prover.  Hence, even though we increased the number of photons by a factor $1/\eta^2$, the information the verifier gets is still $O(\sqrt{N}\log N)$, since most photons get lost. Of course, if we want to be even more stringent and ensure that the verifier could not get more information even if he replaces the entire circuit with a lossless one, then we can include the factor $1/\eta^2$ and upper bound the information as $O(\sqrt{N}\log N/\eta^2)$.


\section{Discussion}
We have shown that it is possible to verify NP-complete problems using simple linear optics. This is done by reducing instances of any NP-complete problem to a balanced instance of 2-out-of-4 SAT. The solution is encoded into single-photon states in a superposition across many optical modes, which are then verified by choosing randomly between three different tests. Each of these tests can be implemented using simple linear optics, namely mode permutations and interferometry of at most four modes. We have also shown that a quantum advantage can be obtained for the running time of verification algorithms of proofs that reveal a limited amount of information about the variables. This advantage only holds if the states are not entangled with each other, which we take as a promise from the prover. Overall, our results provide another example of the surprising computational power of linear optics.

Besides the advantage that we discuss in this paper, there are other features of this protocol that are appealing. One of them is the low energy expenditure of the scheme. The only energy consumption takes place in the state preparation, which uses only a few photons, and in the permutation of the modes. This is likely to be less resource-intensive than running conventional computers for the same verification. Additionally, besides thinking of the limited information of the proofs as a restriction on the verification, we can view it as a security goal of Merlin who wants to convince Arthur without revealing full information about the solution. These questions are studied in the context of zero-knowledge proofs, where verification is possible without revealing any information. However, those protocols are interactive and often very complex. Our verification scheme provides a simple alternative where only partial information is revealed while requiring no interaction.

In terms of experimental realizations of the verification protocol, as discussed before, technology is currently available to perform arbitrary permutations of large number of modes using microelectromechanical systems, which can be employed to build optical switches capable of permuting as many as 1100 modes \cite{kim20031100}. Since the interferometers act on a small number of modes, the entire verification circuits could in principle be built modularly from small integrated chips, which can be manufactured independently from each other. Finally, although significant progress has been made in preparing high-dimensional single photon states {using integrated photonics} \cite{heilmann2015novel}, it remains a challenge to create many such states independently for interferometric experiments. It is likely, however, that the verification can be performed using coherent states instead of single photons, in which case only a single coherent laser source would be needed which can be subsequently split into the desired modes. Indeed, as shown in Ref.~\cite{arrazola2014QC}, given one single-photon state and a linear-optical measurement on it, replacing the single photon with a coherent state leads to measurement statistics that are equivalent to a randomly-selected number of repetitions of the same measurement on a single-photon state. Further work is needed to ensure that the quantum superiority is retained in this case.

\begin{acknowledgments}
We would like to thank A. Ignjatovic for valuable discussions. J.M.A. acknowledges funding from the Singapore Ministry of Education (partly through the Academic Research Fund Tier 3 MOE2012-T3-1-009) and the National Research Foundation of Singapore, Prime Minister’s Office, under the Research Centres of Excellence programme. This research was supported by the European Research Council project QCC (I.K.). 
\end{acknowledgments}


\begin{thebibliography}{40}
\expandafter\ifx\csname natexlab\endcsname\relax\def\natexlab#1{#1}\fi
\expandafter\ifx\csname bibnamefont\endcsname\relax
  \def\bibnamefont#1{#1}\fi
\expandafter\ifx\csname bibfnamefont\endcsname\relax
  \def\bibfnamefont#1{#1}\fi
\expandafter\ifx\csname citenamefont\endcsname\relax
  \def\citenamefont#1{#1}\fi
\expandafter\ifx\csname url\endcsname\relax
  \def\url#1{\texttt{#1}}\fi
\expandafter\ifx\csname urlprefix\endcsname\relax\def\urlprefix{URL }\fi
\providecommand{\bibinfo}[2]{#2}
\providecommand{\eprint}[2][]{\url{#2}}

\bibitem[{\citenamefont{Knill et~al.}(2001)\citenamefont{Knill, Laflamme, and
  Milburn}}]{knill01a}
\bibinfo{author}{\bibfnamefont{E.}~\bibnamefont{Knill}},
  \bibinfo{author}{\bibfnamefont{R.}~\bibnamefont{Laflamme}}, \bibnamefont{and}
  \bibinfo{author}{\bibfnamefont{G.}~\bibnamefont{Milburn}},
  \bibinfo{journal}{Nature} \textbf{\bibinfo{volume}{409}}, \bibinfo{pages}{46}
  (\bibinfo{year}{2001}).

\bibitem[{\citenamefont{Browne and Rudolph}(2004)}]{browne04suba}
\bibinfo{author}{\bibfnamefont{D.~E.} \bibnamefont{Browne}} \bibnamefont{and}
  \bibinfo{author}{\bibfnamefont{T.}~\bibnamefont{Rudolph}},
  \emph{\bibinfo{title}{Resource-efficient linear optical quantum
  computation}}, \bibinfo{howpublished}{quant-ph/0405157}
  (\bibinfo{year}{2004}).

\bibitem[{\citenamefont{Kok et~al.}(2007)\citenamefont{Kok, Munro, Nemoto,
  Ralph, Dowling, and Milburn}}]{kok2007linear}
\bibinfo{author}{\bibfnamefont{P.}~\bibnamefont{Kok}},
  \bibinfo{author}{\bibfnamefont{W.~J.} \bibnamefont{Munro}},
  \bibinfo{author}{\bibfnamefont{K.}~\bibnamefont{Nemoto}},
  \bibinfo{author}{\bibfnamefont{T.~C.} \bibnamefont{Ralph}},
  \bibinfo{author}{\bibfnamefont{J.~P.} \bibnamefont{Dowling}},
  \bibnamefont{and} \bibinfo{author}{\bibfnamefont{G.~J.}
  \bibnamefont{Milburn}}, \bibinfo{journal}{Rev. Mod. Phys.}
  \textbf{\bibinfo{volume}{79}}, \bibinfo{pages}{135} (\bibinfo{year}{2007}).

\bibitem[{\citenamefont{O'Brien}(2007)}]{o2007optical}
\bibinfo{author}{\bibfnamefont{J.~L.} \bibnamefont{O'Brien}},
  \bibinfo{journal}{Science} \textbf{\bibinfo{volume}{318}},
  \bibinfo{pages}{1567} (\bibinfo{year}{2007}).

\bibitem[{\citenamefont{Giovannetti et~al.}(2006)\citenamefont{Giovannetti,
  Lloyd, and Maccone}}]{giovannetti2006quantum}
\bibinfo{author}{\bibfnamefont{V.}~\bibnamefont{Giovannetti}},
  \bibinfo{author}{\bibfnamefont{S.}~\bibnamefont{Lloyd}}, \bibnamefont{and}
  \bibinfo{author}{\bibfnamefont{L.}~\bibnamefont{Maccone}},
  \bibinfo{journal}{Phys. Rev. Lett.} \textbf{\bibinfo{volume}{96}},
  \bibinfo{pages}{010401} (\bibinfo{year}{2006}).

\bibitem[{\citenamefont{Giovannetti et~al.}(2011)\citenamefont{Giovannetti,
  Lloyd, and Maccone}}]{giovannetti2011advances}
\bibinfo{author}{\bibfnamefont{V.}~\bibnamefont{Giovannetti}},
  \bibinfo{author}{\bibfnamefont{S.}~\bibnamefont{Lloyd}}, \bibnamefont{and}
  \bibinfo{author}{\bibfnamefont{L.}~\bibnamefont{Maccone}},
  \bibinfo{journal}{Nature Photonics} \textbf{\bibinfo{volume}{5}},
  \bibinfo{pages}{222} (\bibinfo{year}{2011}).

\bibitem[{\citenamefont{T{\'o}th and Apellaniz}(2014)}]{toth2014quantum}
\bibinfo{author}{\bibfnamefont{G.}~\bibnamefont{T{\'o}th}} \bibnamefont{and}
  \bibinfo{author}{\bibfnamefont{I.}~\bibnamefont{Apellaniz}},
  \bibinfo{journal}{Journal of Physics A: Mathematical and Theoretical}
  \textbf{\bibinfo{volume}{47}}, \bibinfo{pages}{424006}
  (\bibinfo{year}{2014}).

\bibitem[{\citenamefont{Arrazola and L\"utkenhaus}(2014)}]{arrazolaqfp}
\bibinfo{author}{\bibfnamefont{J.~M.} \bibnamefont{Arrazola}} \bibnamefont{and}
  \bibinfo{author}{\bibfnamefont{N.}~\bibnamefont{L\"utkenhaus}},
  \bibinfo{journal}{Phys. Rev. A} \textbf{\bibinfo{volume}{89}},
  \bibinfo{pages}{062305} (\bibinfo{year}{2014}).

\bibitem[{\citenamefont{Arrazola and L{\"u}tkenhaus}(2014)}]{arrazola2014QC}
\bibinfo{author}{\bibfnamefont{J.~M.} \bibnamefont{Arrazola}} \bibnamefont{and}
  \bibinfo{author}{\bibfnamefont{N.}~\bibnamefont{L{\"u}tkenhaus}},
  \bibinfo{journal}{Phys. Rev. A} \textbf{\bibinfo{volume}{90}},
  \bibinfo{pages}{042335} (\bibinfo{year}{2014}).

\bibitem[{\citenamefont{Xu et~al.}(2015)\citenamefont{Xu, Arrazola, Wei, Wang,
  Palacios-Avila, Feng, Sajeed, L{\"u}tkenhaus, and Lo}}]{xu2015experimental}
\bibinfo{author}{\bibfnamefont{F.}~\bibnamefont{Xu}},
  \bibinfo{author}{\bibfnamefont{J.~M.} \bibnamefont{Arrazola}},
  \bibinfo{author}{\bibfnamefont{K.}~\bibnamefont{Wei}},
  \bibinfo{author}{\bibfnamefont{W.}~\bibnamefont{Wang}},
  \bibinfo{author}{\bibfnamefont{P.}~\bibnamefont{Palacios-Avila}},
  \bibinfo{author}{\bibfnamefont{C.}~\bibnamefont{Feng}},
  \bibinfo{author}{\bibfnamefont{S.}~\bibnamefont{Sajeed}},
  \bibinfo{author}{\bibfnamefont{N.}~\bibnamefont{L{\"u}tkenhaus}},
  \bibnamefont{and} \bibinfo{author}{\bibfnamefont{H.-K.} \bibnamefont{Lo}},
  \bibinfo{journal}{Nature {C}ommunications} \textbf{\bibinfo{volume}{6}},
  \bibinfo{pages}{8735} (\bibinfo{year}{2015}).

\bibitem[{\citenamefont{Guan et~al.}(2016)\citenamefont{Guan, Xu, Yin, Li,
  Zhang, Chen, Yang, Li, You, Chen et~al.}}]{guan2016observation}
\bibinfo{author}{\bibfnamefont{J.-Y.} \bibnamefont{Guan}},
  \bibinfo{author}{\bibfnamefont{F.}~\bibnamefont{Xu}},
  \bibinfo{author}{\bibfnamefont{H.-L.} \bibnamefont{Yin}},
  \bibinfo{author}{\bibfnamefont{Y.}~\bibnamefont{Li}},
  \bibinfo{author}{\bibfnamefont{W.-J.} \bibnamefont{Zhang}},
  \bibinfo{author}{\bibfnamefont{S.-J.} \bibnamefont{Chen}},
  \bibinfo{author}{\bibfnamefont{X.-Y.} \bibnamefont{Yang}},
  \bibinfo{author}{\bibfnamefont{L.}~\bibnamefont{Li}},
  \bibinfo{author}{\bibfnamefont{L.-X.} \bibnamefont{You}},
  \bibinfo{author}{\bibfnamefont{T.-Y.} \bibnamefont{Chen}},
  \bibnamefont{et~al.}, \bibinfo{journal}{Phys. Rev. Lett.}
  \textbf{\bibinfo{volume}{116}}, \bibinfo{pages}{240502}
  (\bibinfo{year}{2016}).

\bibitem[{\citenamefont{Kumar et~al.}(2017)\citenamefont{Kumar, Diamanti, and
  Kerenidis}}]{kumar2017efficient}
\bibinfo{author}{\bibfnamefont{N.}~\bibnamefont{Kumar}},
  \bibinfo{author}{\bibfnamefont{E.}~\bibnamefont{Diamanti}}, \bibnamefont{and}
  \bibinfo{author}{\bibfnamefont{I.}~\bibnamefont{Kerenidis}},
  \bibinfo{journal}{Phys. Rev. A} \textbf{\bibinfo{volume}{95}},
  \bibinfo{pages}{032337} (\bibinfo{year}{2017}).

\bibitem[{\citenamefont{Aaronson and
  Arkhipov}(2011)}]{aaronson2011computational}
\bibinfo{author}{\bibfnamefont{S.}~\bibnamefont{Aaronson}} \bibnamefont{and}
  \bibinfo{author}{\bibfnamefont{A.}~\bibnamefont{Arkhipov}}, in
  \emph{\bibinfo{booktitle}{Proceedings of the forty-third annual ACM symposium
  on Theory of computing}} (\bibinfo{organization}{ACM}, \bibinfo{year}{2011}),
  pp. \bibinfo{pages}{333--342}.

\bibitem[{\citenamefont{Broome et~al.}(2013)\citenamefont{Broome, Fedrizzi,
  Rahimi-Keshari, Dove, Aaronson, Ralph, and White}}]{broome2013photonic}
\bibinfo{author}{\bibfnamefont{M.~A.} \bibnamefont{Broome}},
  \bibinfo{author}{\bibfnamefont{A.}~\bibnamefont{Fedrizzi}},
  \bibinfo{author}{\bibfnamefont{S.}~\bibnamefont{Rahimi-Keshari}},
  \bibinfo{author}{\bibfnamefont{J.}~\bibnamefont{Dove}},
  \bibinfo{author}{\bibfnamefont{S.}~\bibnamefont{Aaronson}},
  \bibinfo{author}{\bibfnamefont{T.~C.} \bibnamefont{Ralph}}, \bibnamefont{and}
  \bibinfo{author}{\bibfnamefont{A.~G.} \bibnamefont{White}},
  \bibinfo{journal}{Science} \textbf{\bibinfo{volume}{339}},
  \bibinfo{pages}{794} (\bibinfo{year}{2013}).

\bibitem[{\citenamefont{Tillmann et~al.}(2013)\citenamefont{Tillmann,
  Daki{\'c}, Heilmann, Nolte, Szameit, and Walther}}]{tillmann2013experimental}
\bibinfo{author}{\bibfnamefont{M.}~\bibnamefont{Tillmann}},
  \bibinfo{author}{\bibfnamefont{B.}~\bibnamefont{Daki{\'c}}},
  \bibinfo{author}{\bibfnamefont{R.}~\bibnamefont{Heilmann}},
  \bibinfo{author}{\bibfnamefont{S.}~\bibnamefont{Nolte}},
  \bibinfo{author}{\bibfnamefont{A.}~\bibnamefont{Szameit}}, \bibnamefont{and}
  \bibinfo{author}{\bibfnamefont{P.}~\bibnamefont{Walther}},
  \bibinfo{journal}{Nature Photonics} \textbf{\bibinfo{volume}{7}},
  \bibinfo{pages}{540} (\bibinfo{year}{2013}).

\bibitem[{\citenamefont{Crespi et~al.}(2013)\citenamefont{Crespi, Osellame,
  Ramponi, Brod, Galvao, Spagnolo, Vitelli, Maiorino, Mataloni, and
  Sciarrino}}]{crespi2013integrated}
\bibinfo{author}{\bibfnamefont{A.}~\bibnamefont{Crespi}},
  \bibinfo{author}{\bibfnamefont{R.}~\bibnamefont{Osellame}},
  \bibinfo{author}{\bibfnamefont{R.}~\bibnamefont{Ramponi}},
  \bibinfo{author}{\bibfnamefont{D.~J.} \bibnamefont{Brod}},
  \bibinfo{author}{\bibfnamefont{E.~F.} \bibnamefont{Galvao}},
  \bibinfo{author}{\bibfnamefont{N.}~\bibnamefont{Spagnolo}},
  \bibinfo{author}{\bibfnamefont{C.}~\bibnamefont{Vitelli}},
  \bibinfo{author}{\bibfnamefont{E.}~\bibnamefont{Maiorino}},
  \bibinfo{author}{\bibfnamefont{P.}~\bibnamefont{Mataloni}}, \bibnamefont{and}
  \bibinfo{author}{\bibfnamefont{F.}~\bibnamefont{Sciarrino}},
  \bibinfo{journal}{Nature Photonics} \textbf{\bibinfo{volume}{7}},
  \bibinfo{pages}{545} (\bibinfo{year}{2013}).

\bibitem[{\citenamefont{Spring et~al.}(2013)\citenamefont{Spring, Metcalf,
  Humphreys, Kolthammer, Jin, Barbieri, Datta, Thomas-Peter, Langford, Kundys
  et~al.}}]{spring2012boson}
\bibinfo{author}{\bibfnamefont{J.~B.} \bibnamefont{Spring}},
  \bibinfo{author}{\bibfnamefont{B.~J.} \bibnamefont{Metcalf}},
  \bibinfo{author}{\bibfnamefont{P.~C.} \bibnamefont{Humphreys}},
  \bibinfo{author}{\bibfnamefont{W.~S.} \bibnamefont{Kolthammer}},
  \bibinfo{author}{\bibfnamefont{X.-M.} \bibnamefont{Jin}},
  \bibinfo{author}{\bibfnamefont{M.}~\bibnamefont{Barbieri}},
  \bibinfo{author}{\bibfnamefont{A.}~\bibnamefont{Datta}},
  \bibinfo{author}{\bibfnamefont{N.}~\bibnamefont{Thomas-Peter}},
  \bibinfo{author}{\bibfnamefont{N.~K.} \bibnamefont{Langford}},
  \bibinfo{author}{\bibfnamefont{D.}~\bibnamefont{Kundys}},
  \bibnamefont{et~al.}, \bibinfo{journal}{Science}
  \textbf{\bibinfo{volume}{339}}, \bibinfo{pages}{798} (\bibinfo{year}{2013}).

\bibitem[{\citenamefont{Spagnolo et~al.}(2014)\citenamefont{Spagnolo, Vitelli,
  Bentivegna, Brod, Crespi, Flamini, Giacomini, Milani, Ramponi, Mataloni
  et~al.}}]{spagnolo2014experimental}
\bibinfo{author}{\bibfnamefont{N.}~\bibnamefont{Spagnolo}},
  \bibinfo{author}{\bibfnamefont{C.}~\bibnamefont{Vitelli}},
  \bibinfo{author}{\bibfnamefont{M.}~\bibnamefont{Bentivegna}},
  \bibinfo{author}{\bibfnamefont{D.~J.} \bibnamefont{Brod}},
  \bibinfo{author}{\bibfnamefont{A.}~\bibnamefont{Crespi}},
  \bibinfo{author}{\bibfnamefont{F.}~\bibnamefont{Flamini}},
  \bibinfo{author}{\bibfnamefont{S.}~\bibnamefont{Giacomini}},
  \bibinfo{author}{\bibfnamefont{G.}~\bibnamefont{Milani}},
  \bibinfo{author}{\bibfnamefont{R.}~\bibnamefont{Ramponi}},
  \bibinfo{author}{\bibfnamefont{P.}~\bibnamefont{Mataloni}},
  \bibnamefont{et~al.}, \bibinfo{journal}{Nature Photonics}
  \textbf{\bibinfo{volume}{8}}, \bibinfo{pages}{615} (\bibinfo{year}{2014}).

\bibitem[{\citenamefont{Boixo et~al.}(2016)\citenamefont{Boixo, Isakov,
  Smelyanskiy, Babbush, Ding, Jiang, Martinis, and
  Neven}}]{boixo2016characterizing}
\bibinfo{author}{\bibfnamefont{S.}~\bibnamefont{Boixo}},
  \bibinfo{author}{\bibfnamefont{S.~V.} \bibnamefont{Isakov}},
  \bibinfo{author}{\bibfnamefont{V.~N.} \bibnamefont{Smelyanskiy}},
  \bibinfo{author}{\bibfnamefont{R.}~\bibnamefont{Babbush}},
  \bibinfo{author}{\bibfnamefont{N.}~\bibnamefont{Ding}},
  \bibinfo{author}{\bibfnamefont{Z.}~\bibnamefont{Jiang}},
  \bibinfo{author}{\bibfnamefont{J.~M.} \bibnamefont{Martinis}},
  \bibnamefont{and} \bibinfo{author}{\bibfnamefont{H.}~\bibnamefont{Neven}},
  \bibinfo{journal}{arXiv preprint arXiv:1608.00263}  (\bibinfo{year}{2016}).

\bibitem[{\citenamefont{Farhi and Harrow}(2016)}]{farhi2016quantum}
\bibinfo{author}{\bibfnamefont{E.}~\bibnamefont{Farhi}} \bibnamefont{and}
  \bibinfo{author}{\bibfnamefont{A.~W.} \bibnamefont{Harrow}},
  \bibinfo{journal}{arXiv preprint arXiv:1602.07674}  (\bibinfo{year}{2016}).

\bibitem[{\citenamefont{Bremner et~al.}(2017)\citenamefont{Bremner, Montanaro,
  and Shepherd}}]{bremner2016achieving}
\bibinfo{author}{\bibfnamefont{M.~J.} \bibnamefont{Bremner}},
  \bibinfo{author}{\bibfnamefont{A.}~\bibnamefont{Montanaro}},
  \bibnamefont{and} \bibinfo{author}{\bibfnamefont{D.~J.}
  \bibnamefont{Shepherd}}, \bibinfo{journal}{Quantum}
  \textbf{\bibinfo{volume}{1}}, \bibinfo{pages}{8} (\bibinfo{year}{2017}).

\bibitem[{\citenamefont{Bravyi et~al.}(2017)\citenamefont{Bravyi, Gosset, and
  Koenig}}]{bravyi2017quantum}
\bibinfo{author}{\bibfnamefont{S.}~\bibnamefont{Bravyi}},
  \bibinfo{author}{\bibfnamefont{D.}~\bibnamefont{Gosset}}, \bibnamefont{and}
  \bibinfo{author}{\bibfnamefont{R.}~\bibnamefont{Koenig}},
  \bibinfo{journal}{arXiv preprint arXiv:1704.00690}  (\bibinfo{year}{2017}).

\bibitem[{\citenamefont{Gao et~al.}(2017)\citenamefont{Gao, Wang, and
  Duan}}]{gao2017quantum}
\bibinfo{author}{\bibfnamefont{X.}~\bibnamefont{Gao}},
  \bibinfo{author}{\bibfnamefont{S.-T.} \bibnamefont{Wang}}, \bibnamefont{and}
  \bibinfo{author}{\bibfnamefont{L.-M.} \bibnamefont{Duan}},
  \bibinfo{journal}{Phys. Rev. Lett.} \textbf{\bibinfo{volume}{118}},
  \bibinfo{pages}{040502} (\bibinfo{year}{2017}).

\bibitem[{\citenamefont{Bermejo-Vega et~al.}(2017)\citenamefont{Bermejo-Vega,
  Hangleiter, Schwarz, Raussendorf, and Eisert}}]{bermejo2017architectures}
\bibinfo{author}{\bibfnamefont{J.}~\bibnamefont{Bermejo-Vega}},
  \bibinfo{author}{\bibfnamefont{D.}~\bibnamefont{Hangleiter}},
  \bibinfo{author}{\bibfnamefont{M.}~\bibnamefont{Schwarz}},
  \bibinfo{author}{\bibfnamefont{R.}~\bibnamefont{Raussendorf}},
  \bibnamefont{and} \bibinfo{author}{\bibfnamefont{J.}~\bibnamefont{Eisert}},
  \bibinfo{journal}{arXiv preprint arXiv:1703.00466}  (\bibinfo{year}{2017}).

\bibitem[{\citenamefont{Clifford and Clifford}(2017)}]{clifford2017classical}
\bibinfo{author}{\bibfnamefont{P.}~\bibnamefont{Clifford}} \bibnamefont{and}
  \bibinfo{author}{\bibfnamefont{R.}~\bibnamefont{Clifford}},
  \bibinfo{journal}{arXiv preprint arXiv:1706.01260}  (\bibinfo{year}{2017}).

\bibitem[{\citenamefont{Neville et~al.}(2017)\citenamefont{Neville, Sparrow,
  Clifford, Johnston, Birchall, Montanaro, and Laing}}]{neville2017no}
\bibinfo{author}{\bibfnamefont{A.}~\bibnamefont{Neville}},
  \bibinfo{author}{\bibfnamefont{C.}~\bibnamefont{Sparrow}},
  \bibinfo{author}{\bibfnamefont{R.}~\bibnamefont{Clifford}},
  \bibinfo{author}{\bibfnamefont{E.}~\bibnamefont{Johnston}},
  \bibinfo{author}{\bibfnamefont{P.~M.} \bibnamefont{Birchall}},
  \bibinfo{author}{\bibfnamefont{A.}~\bibnamefont{Montanaro}},
  \bibnamefont{and} \bibinfo{author}{\bibfnamefont{A.}~\bibnamefont{Laing}},
  \bibinfo{journal}{Nature Physics}  (\bibinfo{year}{2017}),
  \bibinfo{note}{doi:10.1038/nphys4270}.

\bibitem[{\citenamefont{Aaronson et~al.}(2008)\citenamefont{Aaronson, Beigi,
  Drucker, Fefferman, and Shor}}]{unentanglement}
\bibinfo{author}{\bibfnamefont{S.}~\bibnamefont{Aaronson}},
  \bibinfo{author}{\bibfnamefont{S.}~\bibnamefont{Beigi}},
  \bibinfo{author}{\bibfnamefont{A.}~\bibnamefont{Drucker}},
  \bibinfo{author}{\bibfnamefont{B.}~\bibnamefont{Fefferman}},
  \bibnamefont{and} \bibinfo{author}{\bibfnamefont{P.}~\bibnamefont{Shor}}, in
  \emph{\bibinfo{booktitle}{23rd Annual IEEE Conference on Computational
  Complexity, 2008.}} (\bibinfo{organization}{IEEE}, \bibinfo{year}{2008}), pp.
  \bibinfo{pages}{223--236}.

\bibitem[{\citenamefont{Impagliazzo and Paturi}(1999)}]{impagliazzo99}
\bibinfo{author}{\bibfnamefont{R.}~\bibnamefont{Impagliazzo}} \bibnamefont{and}
  \bibinfo{author}{\bibfnamefont{R.}~\bibnamefont{Paturi}},
  \bibinfo{journal}{Proc. 14th IEEE Conf. on Computational Complexity} p.
  \bibinfo{pages}{237–240} (\bibinfo{year}{1999}).

\bibitem[{\citenamefont{Dantsin and Wolpert}(2010)}]{dantsin10}
\bibinfo{author}{\bibfnamefont{E.}~\bibnamefont{Dantsin}} \bibnamefont{and}
  \bibinfo{author}{\bibfnamefont{A.}~\bibnamefont{Wolpert}},
  \bibinfo{journal}{Theory and Applications of Satisfiability Testing –- SAT}
  pp. \bibinfo{pages}{313--325} (\bibinfo{year}{2010}).

\bibitem[{\citenamefont{Goldwasser et~al.}(1989)\citenamefont{Goldwasser,
  Micali, and Rackoff}}]{goldwasser89}
\bibinfo{author}{\bibfnamefont{S.}~\bibnamefont{Goldwasser}},
  \bibinfo{author}{\bibfnamefont{S.}~\bibnamefont{Micali}}, \bibnamefont{and}
  \bibinfo{author}{\bibfnamefont{C.}~\bibnamefont{Rackoff}},
  \bibinfo{journal}{SIAM Journal on Computing}  (\bibinfo{year}{1989}).

\bibitem[{\citenamefont{Irit}(2007)}]{dinur}
\bibinfo{author}{\bibfnamefont{D.}~\bibnamefont{Irit}}, \bibinfo{journal}{J.
  ACM} \textbf{\bibinfo{volume}{54(3):12}} (\bibinfo{year}{2007}).

\bibitem[{\citenamefont{Harrow and Montanaro}(2013)}]{2provers}
\bibinfo{author}{\bibfnamefont{A.}~\bibnamefont{Harrow}} \bibnamefont{and}
  \bibinfo{author}{\bibfnamefont{A.}~\bibnamefont{Montanaro}},
  \bibinfo{journal}{J. ACM} \textbf{\bibinfo{volume}{60(1):3}}
  (\bibinfo{year}{2013}).

\bibitem[{\citenamefont{Heilmann et~al.}(2015)\citenamefont{Heilmann,
  Gr{\"a}fe, Nolte, and Szameit}}]{heilmann2015novel}
\bibinfo{author}{\bibfnamefont{R.}~\bibnamefont{Heilmann}},
  \bibinfo{author}{\bibfnamefont{M.}~\bibnamefont{Gr{\"a}fe}},
  \bibinfo{author}{\bibfnamefont{S.}~\bibnamefont{Nolte}}, \bibnamefont{and}
  \bibinfo{author}{\bibfnamefont{A.}~\bibnamefont{Szameit}},
  \bibinfo{journal}{Science} \textbf{\bibinfo{volume}{60}}, \bibinfo{pages}{96}
  (\bibinfo{year}{2015}).

\bibitem[{\citenamefont{Knight and Bloom}(1973)}]{knight1973birthday}
\bibinfo{author}{\bibfnamefont{W.}~\bibnamefont{Knight}} \bibnamefont{and}
  \bibinfo{author}{\bibfnamefont{D.}~\bibnamefont{Bloom}},
  \bibinfo{journal}{American Mathematical Monthly}
  \textbf{\bibinfo{volume}{80}}, \bibinfo{pages}{1141} (\bibinfo{year}{1973}).

\bibitem[{\citenamefont{Hong et~al.}(1987)\citenamefont{Hong, Ou, and
  Mandel}}]{hong1987measurement}
\bibinfo{author}{\bibfnamefont{C. K.}~\bibnamefont{Hong}},
  \bibinfo{author}{\bibfnamefont{Z. Y.}~\bibnamefont{Ou}}, \bibnamefont{and}
  \bibinfo{author}{\bibfnamefont{L.}~\bibnamefont{Mandel}},
  \bibinfo{journal}{Phys. Rev. Lett.} \textbf{\bibinfo{volume}{59}},
  \bibinfo{pages}{2044} (\bibinfo{year}{1987}).

\bibitem[{\citenamefont{Garcia-Escartin and
  Chamorro-Posada}(2013)}]{garcia2013swap}
\bibinfo{author}{\bibfnamefont{J.~C.} \bibnamefont{Garcia-Escartin}}
  \bibnamefont{and}
  \bibinfo{author}{\bibfnamefont{P.}~\bibnamefont{Chamorro-Posada}},
  \bibinfo{journal}{Phys. Rev. A} \textbf{\bibinfo{volume}{87}},
  \bibinfo{pages}{052330} (\bibinfo{year}{2013}).

\bibitem[{\citenamefont{Buhrman et~al.}(2001)\citenamefont{Buhrman, Cleve,
  Watrous, and de~Wolf}}]{QuantumFingerprinting}
\bibinfo{author}{\bibfnamefont{H.}~\bibnamefont{Buhrman}},
  \bibinfo{author}{\bibfnamefont{R.}~\bibnamefont{Cleve}},
  \bibinfo{author}{\bibfnamefont{J.}~\bibnamefont{Watrous}}, \bibnamefont{and}
  \bibinfo{author}{\bibfnamefont{R.}~\bibnamefont{de~Wolf}},
  \bibinfo{journal}{Phys. Rev. Lett.} \textbf{\bibinfo{volume}{87}},
  \bibinfo{pages}{167902} (\bibinfo{year}{2001}).

\bibitem[{\citenamefont{Reck et~al.}(1994)\citenamefont{Reck, Zeilinger,
  Bernstein, and Bertani}}]{reck94a}
\bibinfo{author}{\bibfnamefont{M.}~\bibnamefont{Reck}},
  \bibinfo{author}{\bibfnamefont{A.}~\bibnamefont{Zeilinger}},
  \bibinfo{author}{\bibfnamefont{H.~J.} \bibnamefont{Bernstein}},
  \bibnamefont{and} \bibinfo{author}{\bibfnamefont{P.}~\bibnamefont{Bertani}},
  \bibinfo{journal}{Phys. Rev. Lett.} \textbf{\bibinfo{volume}{73}},
  \bibinfo{pages}{58} (\bibinfo{year}{1994}).

\bibitem[{\citenamefont{Carolan et~al.}(2015)\citenamefont{Carolan, Harrold,
  Sparrow, Martin-Lopez, Russell, Silverstone, Shadbolt, Matsuda, Oguma, Itoh
  et~al.}}]{carolan2015universal}
\bibinfo{author}{\bibfnamefont{J.}~\bibnamefont{Carolan}},
  \bibinfo{author}{\bibfnamefont{C.}~\bibnamefont{Harrold}},
  \bibinfo{author}{\bibfnamefont{C.}~\bibnamefont{Sparrow}},
  \bibinfo{author}{\bibfnamefont{E.}~\bibnamefont{Martin-Lopez}},
  \bibinfo{author}{\bibfnamefont{N.~J.} \bibnamefont{Russell}},
  \bibinfo{author}{\bibfnamefont{J.~W.} \bibnamefont{Silverstone}},
  \bibinfo{author}{\bibfnamefont{P.~J.} \bibnamefont{Shadbolt}},
  \bibinfo{author}{\bibfnamefont{N.}~\bibnamefont{Matsuda}},
  \bibinfo{author}{\bibfnamefont{M.}~\bibnamefont{Oguma}},
  \bibinfo{author}{\bibfnamefont{M.}~\bibnamefont{Itoh}}, \bibnamefont{et~al.},
  \bibinfo{journal}{Science} \textbf{\bibinfo{volume}{349}},
  \bibinfo{pages}{711} (\bibinfo{year}{2015}).

\bibitem[{\citenamefont{Kim et~al.}(2003)\citenamefont{Kim, Nuzman, Kumar,
  Lieuwen, Kraus, Weiss, Lichtenwalner, Papazian, Frahm, Basavanhally
  et~al.}}]{kim20031100}
\bibinfo{author}{\bibfnamefont{J.}~\bibnamefont{Kim}},
  \bibinfo{author}{\bibfnamefont{C.}~\bibnamefont{Nuzman}},
  \bibinfo{author}{\bibfnamefont{B.}~\bibnamefont{Kumar}},
  \bibinfo{author}{\bibfnamefont{D.}~\bibnamefont{Lieuwen}},
  \bibinfo{author}{\bibfnamefont{J.}~\bibnamefont{Kraus}},
  \bibinfo{author}{\bibfnamefont{A.}~\bibnamefont{Weiss}},
  \bibinfo{author}{\bibfnamefont{C.}~\bibnamefont{Lichtenwalner}},
  \bibinfo{author}{\bibfnamefont{A.}~\bibnamefont{Papazian}},
  \bibinfo{author}{\bibfnamefont{R.}~\bibnamefont{Frahm}},
  \bibinfo{author}{\bibfnamefont{N.}~\bibnamefont{Basavanhally}},
  \bibnamefont{et~al.}, \bibinfo{journal}{IEEE Photonics Technology Letters}
  \textbf{\bibinfo{volume}{15}}, \bibinfo{pages}{1537} (\bibinfo{year}{2003}).

\end{thebibliography}

\end{document}